\documentclass[aps,prd,floatfix,showpacs]{revtex4}
\usepackage[dvips]{graphicx}
\newcommand{\be}{\begin{equation}}
\newcommand{\ee}{\end{equation}}
\newcommand{\bea}{\begin{eqnarray}}
\newcommand{\eea}{\end{eqnarray}}
\newcommand{\bitem}{\begin{itemize}}
\newcommand{\eitem}{\end{itemize}}

\newcommand{\benum}{\begin{enumerate}}
\newcommand{\eenum}{\end{enumerate}}
\newcommand{\bc}{\begin{center}}
\newcommand{\ec}{\end{center}}
\begin{document}
\title{APPLICATIONS OF CLASSICAL SCALING SYMMETRY}
\author{Sidney Bludman}
\email{sbludman@das.uchile.cl}
\homepage{http://www.das.uchile.cl/~sbludman}
\affiliation{Departamento de Astronom\'ia, Universidad de Chile, Santiago, Chile}
\date{\today}

\begin{abstract}
Any symmetry reduces a second-order differential equation to a first-order equation: variational symmetries of the action (exemplified by central field dynamics) lead to conservation laws, but
symmetries of only the equations of motion (exemplified by scale-invariant hydrostatics), yield first-order {\em non-conservation laws} between invariants.  We obtain these conservation laws by extending Noether's Theorem to non-variational symmetries,  and present a variational formulation of spherical adiabatic hydrostatics.  For scale-invariant hydrostatics, we directly
recover all the published properties of polytropes and define a {\em core radius}, a new measure of mass concentration in polytropes of index $n$. The Emden solutions (regular solutions of the Lane-Emden equation) are finally obtained, along with useful approximations. An appendix discusses
the special $n=3$ polytrope, emphasizing how the same mechanical structure allows different {\em thermostatic} structures in relativistic degenerate white dwarfs and and zero age main sequence stars.

\end{abstract}
\pacs{45.20.Jj, 45.50.-j, 47.10.A-, 47.10.ab, 47.10.Df, 95.30.Lz, 97.10.Cv}
\maketitle
%\tableofcontents
\section{SYMMETRY REDUCES THE ORDER OF A DIFFERENTIAL EQUATION}
Noether's Theorem relates every {\em variational symmetry} to a conservation law, a first integral of the equations of motion, which can then be integrated directly by quadrature \cite{Bluman}. By an extension of Noether's Theorem, non-variational {\em symmetries of the equations of motion} also reduce them to {\em reduced} equations, which are not conservation laws \cite{BludKenI}.
Consider any system described by the Lagrangian
$\mathcal{L}(t,q_i,\dot{q_i})$ and action $S=\int{\mathcal{L}(t,q_i,\dot{q_i})} dt$, where the $q_i$ are the coordinates, the dot designates the partial derivative $\partial/\partial t$ with respect to the independent variable, and the Einstein summation convention is assumed for repeated indices.   Under any infinitesimal point transformation
$\delta (t,q_i), \delta q_j (t,q_i)$ generated by $\delta t \cdot\partial/\partial t+
\delta q_i\cdot\partial/\partial q_i$,
velocities and Lagrangian transform locally as
\bea \delta \dot{q_i}=\frac{d \delta q_i}{d t}-\dot{q_i}\frac{d \delta t}{d t} \quad , \nonumber \\
\delta \mathcal{L}=\dot{\mathcal{L}}\delta t+(\partial\mathcal{L}/\partial q_i)\delta q_i+(\partial\mathcal{L}/\partial\dot{q_i})
\Bigl[\frac{d \delta q_i}{d t}-\dot{q_i}\frac{d \delta t}{d t}\Bigr]=\Bigl
[\frac{dG}{dt}-
 \mathcal{L}\cdot\frac{d(\delta t)}{dt}+\mathcal{D}_i\cdot(\delta q_i-\dot{q_i}\delta t)\Bigr] \quad,
\eea
in terms of the total derivative of the generator of the transformation, the {\em Noether charge}
\be G:=\mathcal{L}\cdot\delta t+p_i\cdot(\delta q_i-\dot{q_i}\delta t) \quad,\ee
and the Euler-Lagrange variational derivative $\mathcal{D}_i:=
  \partial\mathcal{L}/\partial q_i-d(\partial\mathcal{L}/\partial\dot{q_i})/dt$.
The variation in action between fixed end points is
\be \delta S_{12}=\int_2^1 dt\ \delta\mathcal{L}=\int_2^1 dt\ \Bigl[\frac{dG}{dt}-
 \mathcal{L}\cdot\frac{d(\delta t)}{dt}+\mathcal{D}_i\cdot(\delta q_i-\dot{q_i}\delta t)\Bigr]=
  G(1)-G(2)+\int_2^1 dt\ \Bigl[\delta q_i\cdot\mathcal{D}_i+\delta t\cdot \Bigl(\frac{d \mathcal{H}}{dt}+
  \frac{\partial\mathcal{L}}{\partial t}\Bigr) \Bigr]\quad , \ee
after integrating the term in $d(\delta t)/dt$ by parts.
The action principle asserts that this variation vanishes for independent variations $\delta q_i, \delta t$ that vanish at the end points.  It implies the Euler-Lagrange equations $\mathcal{D}_i=0$ and $d \mathcal{H}/dt=-\partial\mathcal{L}/\partial t$, the rate of change of the Hamiltonian in non-holonomic systems.

On-shell, where $\mathcal{D}_i=0$,
\bea \delta S_{12}=\int_2^1{ \bar{\delta}\mathcal{L}}\ dt=G(1)-G(2) \\
\frac{dG}{dt} =\bar{\delta}\mathcal{L}:=\delta\mathcal{L}+\mathcal{L}\cdot (d\delta t/dt) . \eea
This extension of Noether's Theorem describes the evolution of {\em any} generator or Noether charge, in terms of the transformation of the Lagrangian it generates. It expresses the Euler-Lagrange equations of motion as the divergence of a Noether charge, which vanishes for a variational symmetry, but not for any other symmetry transformation (Section I).

Our primary purpose is to contrast these different ways of reducing second-order differential equations to first-order, by comparing two familiar physical examples:
\begin{description}
\item[Central Field Motion in a Static Potential] which is completely integrable by virtue of energy and angular momentum conservation, whether or not the system is scale invariant (Section II).
\item[Hydrostatic Gaseous Spheres in Adiabatic Equilibrium] which are integrable only if they are scale invariant (polytropes) (Section III).
\end{description}

Although  generally not a conservation law, any symmetry of the equations of motion implies a useful dynamical or structural first-order equation \cite{Bluman}.
Scaling symmetry is the most general simplification that one can make for any dynamical system.  For the radial scaling transformations we will consider, $\delta r=r$, the Lagrangian scales as some scalar density $\delta \mathcal{L}=-2\tilde{\omega}\mathcal{L}$ and the action scales as $\delta S=(1-2\tilde{\omega}) S$.
The Noether charge generating the scale transformation evolves according to a {\em non-conservation law} $\frac{dG}{dt} =(1-2\tilde{\omega})\mathcal{L}$ \cite{BludKenI}, a first-order equation encapsulating all of the consequences of scaling symmetry.
All the published properties
\cite{Chandra,Hansen,Kippen}
 of index-n polytropes follow directly from this first-order equation.

 Our secondary purpose is to present our original variational formulation of spherical hydrostatics, our {\em direct} application of the scaling non-conservation law deriving from our extension of Noether's Theorem to non-variational scaling symmetry, our definition of a {\em core radius}, inside which all polytropes exhibit a common mass density structure (Sections III, IV). Section V completes the integration of the Lane-Emden equation by quadratures and obtains useful approximations to the Emden function $\theta_n (\xi)$.

An appendix reviews the thermodynamic properties of the physically important polytropes of index $n=3$ \cite{Hansen,Kippen,BludKenI}. What is original here is the explanation of the
the differences
between relativistic degenerate white dwarf stars and ideal gas stars on the zero-age main sequence (ZAMS), following from their different entropy structures. Our original approximations to $\theta_3 (\xi)$ should prove useful in such stars.
\section{VARIATIONAL SYMMETRIES IMPLY CONSERVATION LAWS} %%II p2
Time translation and spatial rotations are variational symmetries of the action integral,
\be S=\int\mathcal{L} (r,\dot{r},\dot{\theta})~dt \quad,\quad \mathcal{L}:=T(r,\dot{r},\dot{\theta})-V(r)=\frac{m}{2}(\dot{r}^2+r^2\dot{\theta}^2)-V(r)\quad,\quad \ee
so that
the energy and angular momentum
\be E (r, \dot{r})=(m/2)\mathbf{\dot{r}}^2  +V(r) =(m/2)(\dot{r} ^2+(l/mr)^2)+ V(r)
  \quad ,\qquad l=m r^2 \dot{\theta} , \label{eq:centralforce} \ee
are conserved. Because of these two first integrals, conservative central field motion is completely integrable by quadrature
\bea \theta(r)=\theta_0 + \int_{r_0}^r dr/\bigl\lbrace \sqrt{2 m r^4 [E-V(r)]/l^2-r^2}\bigr\rbrace\\
           t(r)=t_0+\int_{r_0}^r dr/\sqrt{2 r^2 [E-V(r)]-(l/m)^2} ,
\eea
where $\theta_0,~r_0$ are initial values at time $t_0$.

These two first integrals imply the first-order differential equations
\be \dot{\theta}=l/mr^2 , \quad \dot{r}=\sqrt{\frac{2}{m}[E-V(r)]-(\frac{l}{mr})^2} . \ee
and the \emph{orbit equation}
\be  \frac{dr}{d\theta}=\sqrt{2 m r^4 [E-V(r)]/l^2-r^2} .\ee

\begin{table}[t] %%%%Table I p2
\caption {Period-Amplitude Relations and Virial Theorems for Inverse Power-Law Potentials $V \sim 1/r^n$}
\begin{tabular}{|l|l|l|l|}
\hline\hline
$n$&System&Period-amplitude relation $t\sim r^{1+n/2}$&Virial theorem\\
\hline\hline
-2 & isotropic harmonic oscillator             & period independent of amplitude& $\langle K\rangle =\langle
V\rangle$ \\
-1 & uniform gravitational field                      & falling from rest, e.g., $z=gt^2 /2$  & $\langle
K\rangle =\langle V\rangle /2$ \\
0  & free particles                                 & constant velocity $r\sim t $          & $\langle
K\rangle =0$                \\
1  & Newtonian potential            & Kepler's Third Law $t^2\sim r^3$      & $\langle K\rangle =-\langle
V\rangle /2$ \\
2  & inverse-cube force             & $t \sim r^2 $                         & $\langle K\rangle =-\langle
V\rangle $ \\
\hline\hline
\end{tabular}
\end{table}

What additional consequences follow if, $V(r) \sim 1/r^n$, so that
the equations of motion are also invariant under the {\em infinitesimal scale transformation},
\be \delta t=(1+n/2) t\quad , \quad \delta\mathbf{r}=\mathbf{r},\quad \delta\dot{r}=(-n/2)\dot{r} ,\quad \label{eq:variasym} \ee
which is not a variational symmetry of the action?
Instead of another conservation law, scale invariance implies $t/ r^{(1+n/2)}=constant$ and the period-amplitude relations in Table I. Because the kinetic and potential energies transform infinitesimally as
\be \delta K=-n K\quad ,\quad \delta V=-n V ,\ee
the time derivative of the {\em virial} $A:=\sum \mathbf{p}_i \cdot \mathbf{r}_i$ of a many-body system obeys
\be \dot{A}=2K+n V \quad. \ee
In a bounded system its time average $<\dot{A}=0>$, so that the time averages $<K>,~<V>$ obey the generalized virial theorems in the last column of Table I \cite{BludKenI}.

\section{SCALING SYMMETRY MAKES HYDROSTATIC STELLAR STRUCTURE INTEGRABLE}  %%%IIIp3
\subsection{Variational Principle for Spherical Hydrostatics}
A non-rotating gaseous sphere in hydrostatic equilibrium obeys the equations of hydrostatic equilibrium and mass continuity
\be -d P/\rho dr= G m/r^2 , \quad d m/d r=4\pi r^2 \rho \quad, \label{eq:masspressbalance} \ee
where the pressure, mass density, and included mass $P(r),~\rho(r),~m(r)$ depend on radius $r$. As dependent variables, we prefer to use the gravitational potential $V(r)=\int_\infty^r Gm/r^2 dr$ and specific enthalpy (ejection energy, thermostatic potential) $H(r)=\int_{0}^{P(r)} dP/\rho $, so that (\ref{eq:masspressbalance}) and its integrated form 
\be -d H/d r=d V/d r ,\quad V(r)+H(r)=-\frac{GM}{R}\quad, \ee
express  the conservation of the specific energy as sum of gravitational and internal energies, in a star of mass $M$ and radius $R$.
In terms of the enthalpy $H(r)$, these two first-order equations (\ref{eq:masspressbalance}) are equivalent to the second-order equation of hydrostatic equilibrium (\ref{eq:secondorder}) 
\be \frac{1}{r^2}\frac{d}{dr} \Big( r^2\frac{d H}{dr} \Big) + 4 \pi G \rho(H)=0 \quad , \label{eq:secondorder} \ee
which is Poisson's Law for the gravitational potential $V(r)=-H(r)-\frac{GM}{R}$.

Because $\rho(r),~P(r),~H(r)$ are even functions of the radius $r$,
at the origin to order $r^2$, spherical symmetry $d P/dr=0$ and mass continuity requires, 
\be \rho(r)=\rho_c (1-Ar^2), ~~m(r)=\frac{4 \pi r^3}{3}\cdot(1-\frac{3}{5}A r^2)=\frac{4 \pi r^3}{3} \cdot
\rho_c^{2/5} \rho(r) ^{3/5} \quad . \label{eq:origin} \ee
Thus, the average mass density inside radius $r$ is $\bar{\rho}(r):=\frac{m(r)}{4\pi r^3/3}=\rho_c^{2/5} \rho(r) ^{3/5}$.

The second-order equation of hydrostatic equilibrium (\ref{eq:secondorder})
follows from the variational principle $\delta W=0$ minimizing the  Gibbs free energy
\be W:=\int_0^R dr \mathcal{L}(r,H,H') \quad  \ee \cite{BludKenI} .
The Lagrangian
\be \mathcal{L}(r,H,H')=4\pi r^2[-H'^2/8 \pi G+P(\rho)] \quad ,\quad ':=d /dr\quad ,\ee
is the sum of the gravitational and internal specific energies per radial shell $d r$.
The canonical momentum and Hamiltonian are
\be m:=\partial\mathcal{L}/\partial H'=-r^2 H'/G\quad ,\quad
\mathcal{H}(r,H,m)=-Gm^2/2 r^2-4\pi r^2 P(H)\quad  \label{eq:hydrostaticHamilton} \ee
are the included mass and energy per mass shell.
The canonical equations are
\be \partial\mathcal{H}/\partial m=H'=-Gm/r^2\quad ,
\quad \partial\mathcal{H}/\partial H=-m'=-4\pi r^2\rho \quad .\ee
Spherical geometry makes the system nonautonomous, so that $\partial\mathcal{H}/\partial r=-\partial\mathcal{L}/\partial r=-2 \mathcal{L}/r$
vanishes only asymptotically, as the mass shells approach planarity.

\subsection{First-Order Equation Between Scale Invariants}  %%%p4b

The equations of hydrostatic equilibrium (15) can always be rewritten
\be d\log{u}/d\log{r}=3-u(r)-w(r)\quad ,\quad
 d\log{w}/d\log{r}=u(r)-1+w(r)/n(r)-d\log{[1+n(r)]}/d\log{r}\quad, \label{eq:equil} \ee
in terms of the logarithmic derivatives
\be u(r):= d\log{m}/d\log{r}, \quad v(r):=-d\log{(P/\rho)}/d\log{r}, \quad w(r):=n(r) v(r)= -d\log{\rho}/d\log{r}\quad,\ee
and an index $n(r)$
\be n(r):=d\log{\rho}/d \log{(P/\rho)} \quad,\quad 1+\frac{1}{n (r)}:=d\log{P} /d\log{\rho}\quad ,\ee
which depends on the local thermal structure. Our homology mass density invariant $w(r)$ will make explicit the universal mass density structure of all stellar cores, which is not apparent in the conventional pressure invariant $v_n$.

Because stars never have uniform mass density ($n(r)\neq 0$), their action cannot be invariant under radial translation. A hydrostatic structure will still be completely integrable, if the structural equations (15) are invariant under the infinitesimal {\em scaling transformation}
\be \delta r=r,\quad \delta\rho=-n\tilde{\omega}_n\rho,\quad\delta H=-\tilde{\omega}_n H,\quad\delta H'=-(1+\tilde{\omega}_n) H',\qquad \tilde{\omega}_n := 2/(n-1) \quad , \ee
generated by the Noether charge
\be G_n:=-\mathcal{H}\cdot r -m\cdot(\tilde{\omega}_n H)= r^2 [ (\frac{H'^2}{2 G}+4\pi P(H))\cdot r+\frac{2 H H'}{G(n-1)} ] .\ee
The Lagrangian (20) then transforms as a scalar density of weight $-2\tilde{\omega}_n$
\be \delta\mathcal{L}=-2\tilde{\omega}_n \mathcal{L} \quad,\quad \delta S_{12}=(1-2\tilde{\omega}_n)\cdot S_{12} \quad, \ee
so that only for $\tilde{\omega}_n=1/2$, the $n=5$ polytrope , is the action invariant and scaling a variational symmetry.

Both structural equations (\ref{eq:equil}) will be autonomous, if and only if $n$ is constant, so that,
$P(r)=K\rho(r)^{1+\frac{1}{n}}$, with the same constant $K$ (related to the entropy) at each radius.  When
this is so,
\bea du/d\log{r}=u(3-u-w_n)\quad ,\quad dw_n/d\log{r}=w_n(u-1+\frac{w_n}{n}) \\
\frac{d\log{w_n}}{ u-1+\frac{w_n}{n} }=\frac{d\log{u}}{(3-u-w_n)}=d\log{r}=\frac{d\log{m}}{u} \label{eq:chareqns12}
\eea
\footnote{These characteristic equations are equivalent to a predator/prey equation in population dynamics ~\cite{Boyce,JordonSmith}. With time $t$ replacing
$-\log{r}$, they are Lotka-Volterra equations, modified by additional spontaneous growth terms $-u^2, ~w_n ^2 /n $ on the right side.
The $uw$ cross-terms lead to growth of the predator $w$ at the expense of the prey $u$, so that a population that is exclusively prey initially ($u=3,~w=0$) is ultimately devoured $u\rightarrow 0$. For the weakest
predator/prey
interaction ($n=5$), the predator takes an infinite time to reach the finite value $w=5$.
For stronger predator/prey interaction ($n<5$), the predator grows infinitely $w\rightarrow\infty$ in
finite time.}.

In this section, we consider only the first equality (30), the first-order equation
\be \frac{dw_n}{du}=\frac{w_n(u-1+w_n /n)}{u(3-u-w _n)} \quad , \label{eq:firstorder} \ee
encapsulating all the effects of scale invariance \cite{BludKenI}. We will consider only polytropes with finite central density $\rho_c$, so that the regularity condition (\ref{eq:origin}) requires that all $w _n(u)$ be tangent to $\frac{5}{3}(3-u)$.
Such {\em Emden polytropes} are
the regular solutions $w_n(u)$ of the first-order equation, for which
$w _n(u)\rightarrow \frac{5}{3}(3-u)$ for $u\rightarrow 3$.
In place of $u$, we now introduce a new homology invariant
$z:=3-u=-d\log{\bar{\rho}_n}/d\log{r}$, where $\bar{\rho}_n:=3 m (r)/4\pi r^3$ is the average mass density inside radius $r$.
\begin{figure}[t]
\includegraphics[scale=0.95]{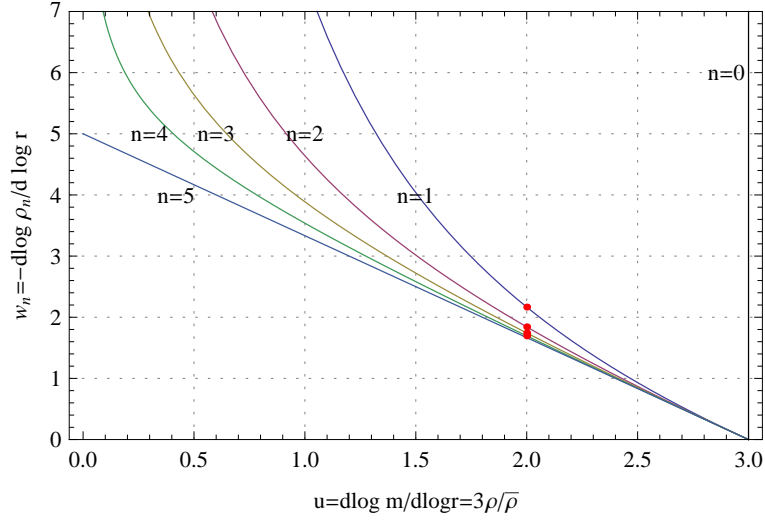} %FIG. 1 p7t
\caption{Polytrope density gradient steepens as the boundary is approached ($u\rightarrow 0$). All solutions are tangent to the same density structure $w_n(z)\rightarrow w_5=(5/3)(3-u)$ at the center ($u=3$), but differ for $u<2$, outside the core radii marked by red dots. Approaching the outer boundary ($u\rightarrow 0$), the density $\rho_n(r)$ falls rapidly, so that its gradient $w_n \rightarrow n[_0\omega_n ^{n-1}/u]^{1/n}$ diverges, for $n<5$.}
\end{figure}
In term of $z,~w_n$, the characteristic differential equations (\ref{eq:chareqns12}) are
\be \frac{d z}{(3-z)(w_n-z)}=\frac{d\log{w_n}}{2-z+\frac{w_n}{n}}=d \log{r}=\frac{d\log{m}}{3-z}\quad .\ee
Incorporating the boundary condition, the first of equations (32) takes the form of a Volterra integral equation ~\cite{BludKen}
\be w_n (z)=\int_0 ^z\ dz\ w_n\frac{(2-z+w_n /n)}{(3-z)(w_n-z)}\approx (5/J_n)[1-(1-z/3)^{J_n}]:=w_{n{\rm Pic}}(z)\quad ,\quad J_n := (9n-10)/(7-n) .\ee
On the right side, the {\em Picard approximation} is defined by inserting the
core values $w_n (z)\approx$
$(5/3)z$ inside the integral. For $n=0,~5$, this Picard approximation is everywhere exact. For intermediate polytropic indices $0<n<5$, the Picard approximation breaks down
approaching the boundary, where $w_n$ diverges as $w_n \rightarrow n[_0\omega_n ^{n-1}/u]^{1/n}$, and is poorest for $n\approx 3$. Figure 1 shows the exact $w_n (u)$ for n=0, 1, 2, 3, 4. 5.

The second-order equation of hydrostatic equilibrium (\ref{eq:secondorder}), takes the dimensionless form of the {\em Lane-Emden equation}
\be \frac{d}{d\xi}\Bigl( \xi^2 \frac{d\theta_n}{d\xi}\Bigr) + \xi^2\theta_n ^n = 0\quad, \label{eq:lane-emden-1} \ee
in terms of the dimensional constant, dimensional radius and dimensional enthalpy
\be \alpha ^2:=\frac{(n+1)}{4\pi G}K\rho_c ^{1/n-1}\quad,\quad \xi:=r/\alpha \quad,\qquad H=H_c\theta_n \quad,\quad \mbox{where} ~~H_c \equiv (n+1) \frac{P_c}{ \rho_c } \equiv (n+1) K \rho^{1/n}_c \quad. \ee
The dimensionless enthalpy is $\theta_n(\xi)$ and the dimensional radius, central density, included mass, mass density, average included mass density, specific gravitational force are
\be r:=\alpha\xi,~\rho_c,~
m(r)=4\pi\rho_c\alpha^3\cdot(-\xi^2\theta_n '),~
\rho_n(r)=\rho_c\cdot\theta_n ^n(\xi),~
\bar{\rho}_n(r):=\frac{m(r)}{4\pi r^3/3}=\rho_c\cdot(-3\theta_n '/\xi),~
g(r)=4\pi\rho_c \alpha^2 (-\theta_n ')\ee
where prime designates the derivative $' :=d/d \xi$. The scale invariants are
\be u=-\xi {\theta_n}^n/\theta_n ' , \qquad v_n = -\xi\theta_n '/\theta_n ,\qquad \omega_n :=(u v_n^n)^{1/(n-1)}=-\xi ^{ 1+\tilde{\omega}_n}\theta_n '  \quad.\ee

The Noether charge
\be G_n(\xi)=-\frac{H_c ^2}{G}\cdot\Bigl\{ \xi^2 \cdot \Bigl[ \xi\Bigl(\frac{\theta_n '^2}{2}+\frac{\theta_n ^{n+1}}{n+1}\Bigr)+(\frac{2}{n-1})\theta_n \theta_n ' \Bigr] \quad,\ee
evolves radially according to
\be \frac{d G_n}{d\xi}=-\frac{H_c ^2}{G}\cdot\frac{d}{d\xi} \Bigl\{ \xi^2 \cdot \Bigl[ \xi\Bigl(\frac{\theta_n '^2}{2}+\frac{\theta_n ^{n+1}}{n+1}\Bigr)+
  (\frac{2}{n-1})\theta_n \theta_n ' \Bigr] \Bigr\}=
  \Bigl( \frac{n-5}{n-1} \Bigr)\cdot(-\frac{H_c ^2}{G}) \cdot\xi^2
  \Bigl( \frac{\theta_n '}{2}-\frac{\theta_n ^{n+1}}{n+1}\Bigr) .\ee
This non-conservation law expresses the radial evolution of energy density per mass shell, from entirely internal ($\theta_n ^{n+1}/(n+1)$) at the center, to entirely gravitational ($\theta_n '^2 /2$) at the stellar surface.

For $n=5$, scaling is a variational symmetry and (39) reduces to a conservation law for the Noether charge
\be G_5=\frac{H_c^2}{G}\cdot \xi^2 [\xi (\frac{\theta_5 '^2}{2}+\frac{\theta_5 ^{6}}{6})+\frac{1}{2}\theta_5 \theta_5 ']=\frac{H_c^2}{G}\cdot    (uv_5 ^3)^{1/2}\cdot[-v_5-u/3 + 1]=constant \quad.\ee
For the Emden solution, $v_5$ is finite at the stellar boundary $u=0$,
the constant vanishes, so that $w_5 (u)=5 v_5 =\frac{5}{3} (3-u)$ everywhere.

For $n<5$, ~$v_n$ diverges at the stellar radius $\xi_1$ , but $\omega_ n\rightarrow {_0\omega_n} $, a finite constant characterizing each Emden function. At the boundary $u=0$,  our density invariant $w_n (u)$ diverges as $n [_0\omega_n ^{n-1} /u]^{1/n}$, and
\be (-\xi ^2 \theta_n ')_1= {_0\omega_n}\cdot \xi_1 ^{\frac{n-3}{n-1}}\quad. \ee

Table II lists these constants $\xi_1, ~_0\omega_n$, along with the global mass
density ratios $\rho_{cn}(R)/\bar\rho_n(R)=( \frac{\xi^3}{3 _0\omega_n} )_1 $ and the ensuing dimensional radius-mass relation $M^{n-1}=[G/(n+1)K]^n \cdot ( 4\pi/{_0\omega_n}^{n-1} ) R^{n-3}$ . Figure 1 shows $w_n(u)$ for $n=0, 1, 2, 3, 4, 5$. Besides the well-known \cite{Chandra,Hansen,Kippen} third, fourth and fifth columns referring to the surface, we have added
the sixth and seventh columns referring to the core radius, to be discussed in Section V.
We have calculated {\em all} of Table II
\emph{directly} from the \emph{first-order} equation (\ref{eq:firstorder}), encapsulating all the effects of scale invariance \cite{BludKenI}
\begin{table*}[t] %%%%Table II p6
\caption{Scaling Exponents, Core Parameters, Surface Parameters, and Mass-Radius Relations for Polytropes of Increasing Mass Concentration. Columns 3-5 are well-known \cite{Chandra,Hansen,Kippen}, but columns 6-7 present a new measure of core concentration.}

\begin{ruledtabular}
\begin{tabular}{|l|l||l|l|l||l|r||r|}
$n$ &$\tilde{\omega}_n$ &$\xi_{1n}$ &$\rho_{cn}(R)/\bar\rho_n(R)$&$_0\omega_n$ &$r_{ncore}/R=\xi_{ncore}/\xi_1$&$m_{ncore}/M$
&Radius-Mass Relation $R^{3-n}\sim M^{1-n}/_0\omega_n$ \\
\hline %\hline
0   &-2         &2.449              &1                  &0.333          &1         &1            &$R\sim M^{1/3}$; mass uniformly distributed \\
1   &$\pm\infty$&3.142              &3.290              &...            &0.66      &0.60         &$R$ independent of $M$ \\
1.5 &4          &3.654              &5.991              &132.4          &0.55      &0.51         &$R\sim M^{-1/3}$\\
2   &2          &4.353              &11.403             &10.50          &0.41      &0.41         &                     \\
3   &1          &6.897              &54.183             &2.018          &0.24      &0.31         &$M$ independent of $R$ \\
4   &2/3        &14.972             &622.408            &0.729          &0.13      &0.24         &                      \\
4.5 &4/7        &31.836             &6189.47            &0.394          &0.08      &0.22         &                 \\
5   &1/2        &$\infty$           &$\infty$           &0              &0         &0.19         &$R=\infty$ for any $M$;mass infinitely concentrated;  \\
\end{tabular}
\end{ruledtabular}
\end{table*}
\section{EMDEN SOLUTIONS AND THEIR APPROXIMATIONS} %%IV p6b
After obtaining $w_n(z):=-d\log{\rho_n}/d\log{r}$, either numerically or by Picard approximation, another integration gives \cite{BludKen}
\bea
\rho_n(z)/\rho_{cn} =\exp{\Bigl\lbrace -\int_0 ^z  \frac{dz\ w_n(z)}{[w_n(z)-z](3-z)} \Bigr\rbrace} \approx(1-z/3)^{5/2} \\
\theta_n =[\rho_n(z)/\rho_{cn}]^{1/n} = \exp{\Bigl\lbrace  -\int_0 ^z  \frac{dz\ w_n(z)}{n [w_n(z)-z](3-z)}\Bigr\rbrace}\approx  (1-z/3)^{5/2n}:=\theta_{n{\rm Pic}} \\
m(z)/M=(\frac{z}{3})^{3/2}\cdot \exp{\Bigl\lbrace \int_3 ^z dz \Bigl\lbrace
\frac{1}{[w_n(z)-z]}-\frac{3}{2z}\Bigr\rbrace \Bigr\rbrace} \approx (\frac{z}{3})^{3/2}\\
r(z)/R=\xi /\xi_{1n}=(\frac{z}{3})^{1/2}\cdot \exp{\Bigl\lbrace  \int_3 ^z  dz \Bigl\lbrace \frac{1}{(3-z)[w_n(z)-z]}-
\frac{1}{2z}\Bigr\rbrace \Bigr\rbrace } \approx \frac{(3z)^{1/2}}{3-z}\quad .  \eea
All the scale dependance now appears in the integration constants $M$ and $R(M)$, which depends on $M$, except for $n=3$.

The Picard approximations
\be \theta_{n{\rm Pic}}(\xi)=(1+\xi^2/6N_n)^{-N_n}\quad, \quad N_n:=5/(3n-5) \ee
to the Emden functions are defined by inserting the core values $w_n (z)\approx$
$(5/3)z$ inside the integral and tabulated
in the last column of Table~III. For polytropic indices $n=0,~5$, this closed form is exact.
For intermediate polytropic indices $0<n<5$, the Picard approximation remains a good approximation through order $\xi^6$, but breaks down approaching the outer boundary. Unfortunately, the Picard approximation is poorest near $n=3$, the astrophysically most important polytrope.
\begin{table*}[t] %%%%Table III p6t
\caption{Taylor Series and Picard Approximations $\theta_{n{\rm Pic}}(\xi)$ to Emden Functions $\theta_{n}(\xi)$}
\begin{ruledtabular}
\begin{tabular}{|l||l||l|l}
$n$  &Emden Function $\theta_n(\xi)$ and Taylor Series &$N_n:=5/(3n-5)$ &Picard Approximation $\theta_{n{\rm Pic}}(\xi):=
(1+\xi^2/6N_n)^{-N_n}$ \\
\hline \hline
0    &$1-\xi^2/6$                                   &-1         &$1-\xi^2/6$                              \\
1    &$\sin{\xi}/\xi=1-\xi^2/6+\xi^4/120-\xi^6/5040+\cdots$&-5/2&$(1-\xi^2 /15)^{5/2}=1-\xi^2/6+\xi^4/120-\xi^6/10800+\cdots$    \\
$n$  &$1-\xi^2/6+n \xi^4/120-n(8n-5)/15120 \xi^6+\cdots$   &$5/(3n-5)$ &$(1+\xi^2/6N_n)^{-N_n}=1-\xi^2/6+n \xi^4/120-n(6n-5) \xi^6 /10800+\cdots$ \\
5    &$(1+\xi^2/3)^{-1/2}$                                 &1/2        &$(1+\xi^2/3)^{-1/2}$ \\
\end{tabular}
\end{ruledtabular}
\end{table*}

This $n=3$ polytrope, which is realized in relativistic degenerate white dwarfs and in the Eddington standard model for luminous zero-age main sequence (ZAMS) stars, is distinguished by a unique $M-R$ relation: the mass $M=(\sqrt{4\pi} /_0\omega_3) (G/4K)^{3/2}$ is independent of radius $R$, depending only on
the constant $K:=P/\rho^{4/3}$.  In these stars, the gravitational and internal energies cancel, making the
total energy $W=\Omega + U=0$, leaving them in neutral mechanical equilibrium at any radius.

Figure 2 compares three approximations to this most important Emden function, shown in yellow, whose Taylor series expansion is
\be \theta_3(\xi)= 1-\xi^2/6+\xi^4 /40-(19/5040) \xi^6+(619/1088640) \xi^8 -(2743/39916800) \xi^{10} + \cdots\quad . \ee

\begin{description}
\item[Tenth-order polynomial approximation] to this Taylor series expansion
\be 1-0.1666667 \xi^2+ 0.025 \xi^4 - 0.0037698 \xi^6 +  0.0005686 \xi^8-0.00006872 \xi^{10} \quad,  \ee
shown in red, diverges badly for $\xi >2.5\approx 1.7~\xi_{3core}$.
\item[Picard approximation]
\bea \theta_{3{\rm Pic}}(\xi)=(1+2\xi^2 /15)^{-5/4}=1-\xi^2 /6+\xi^4 /40 -13\xi^6 /3600 + \cdots \\
=1-0.1666667 \xi^2+0.025 \xi^4-0.003611 \xi^6 +\cdots,
\eea
shown in dashed green, converges and remains a good approximation over the bulk of the star, with $\leq 10\%$ error out to $\xi\approx 3.9$, more than twice the core radius and more than half-way out to the stellar boundary at $\xi_{13}=6.897$. This approximation suffices in white dwarf and ZAMS stars, except for their outer envelopes, which contain little mass and are never polytropic.
Because it satisfies the central boundary condition, but not the outer boundary condition, the Picard approximation underestimates $\theta ' (\xi)$ and overestimates $\theta (\xi)$ outside $\xi \sim 3.9$.
\item[Pad\'{e} rational approximation]~\cite{Seidov}:
\be \theta_{\rm 3Pad}=
\frac{1-\xi^2/108+11 \xi^4/45360}{1+17\xi^2/108+ \xi^4/1008}=1- 0.166667 \xi^2 + 0.025 \xi^4 - 0.00376984 \xi^6 +
 0.0005686 \xi^8 - 0.0000857618 \xi^{10}+\cdots , \ee
shown in dashed heavy black, is a much better and simpler approximation. In fact, this Pad\'{e} approximation is almost exact out to
$\xi_1=6.921$, very close to the outer boundary $\xi_{13}=6.897$.
\end{description}
These simple analytic approximations to $\theta_3 (\xi)$, simplify structural modeling of massive white dwarfs and ZAMS stars.
\begin{figure}[t] %%%%Fig. 2 p10
\includegraphics[scale=0.7]{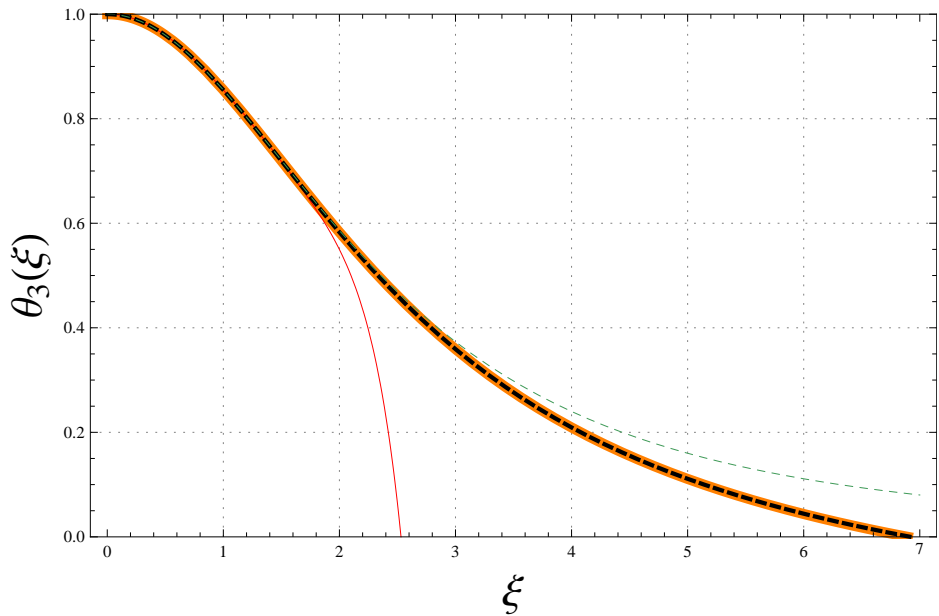}
\caption{The exact Emden function $\theta_3(\xi)$ (solid yellow) and its polynomial (red), Picard (green dashed) and Pad\'{e} (solid black dashed)
approximations. Even in this worst case, the Picard approximation is holds out to twice the core radius
at $2 \xi_{3core}= 3.3$, before breaking down near the boundary. The Pad\'{e} approximation is indistinguishable from the exact solution, vanishing at $\xi_1=6.921$, very close to the exact zero $\xi_{13}=6.897$.}
\end{figure}

\section{A NEW MEASURE FOR CONCENTRATION INCREASING WITH POLYTROPIC INDEX}  %%V

{\em Emden functions} are the normalized solutions of the Lane-Emden equation (\ref{eq:lane-emden-1}) for which the mass density is finite at the origin, so that $\theta_n (0)=1,~\theta_n '(0)=0$.
Each Emden function of index $n$
is characterized by its first zero $\theta_n (\xi_{1n})$, at dimensionless boundary radius $\xi_{1n}$.
As a new measure of core concentration, we also define the {\em core radius} $\xi_{core}$ implicitly by $u(\xi_{core}):=2$,
where gravitational and pressure gradient forces are maximal and
$w_n\approx 2, ~\rho_{ncore}/\rho _{nc}\approx 0.4$ for all polytropes $n\geq 1$.  Inside the core, the specific internal energy density dominates over the gravitational potential, so that for $n\geq 1$,
\be w_n (u)  \approx w_5 (u)=\frac{5}{3} (3-u)\quad,\quad \theta_n (\xi)\approx 1-\xi^2 /6 \quad,\quad \mbox{for}\quad u>2,~\xi<\xi_{core} \label{eq:coredef} \quad ,
 \ee
 consistent with the universal density structure (18) all stars enjoy near their center.
Inside the core, the enthalpy $H(r)\sim\theta_n (\xi)$ decreases as $\exp{(-\xi^2 /6)}$ for all polytropes.  Outside the core, the gravitational potential V(r) dominates as it
increases towards $-GM/R$ at the stellar surface.

The core concentration is illustrated in Figures 3 and 4, which show the local density $\rho_n/\rho_{nc}=\theta_n ^n$ as function of included mass fraction $\frac{m}{M}=\frac{\xi^2
\theta_n '}{(\xi^2 \theta_n ')_1}$ and of fractional radius $\frac{r}{R}=\frac{\xi}{\xi_1}$ respectively.
On each curve $n$ in Figures 1, 3, 4, the core radii is marked by a red dot.

The sixth and seventh columns in Table~II list dimensionless values for the fractional core radius $r_{ncore}/R=\xi_{ncore}/\xi_1$ and fractional included mass $m_{ncore}/M$.
\begin{figure}[t] %%%%Fig. 3
\includegraphics[scale=0.75]{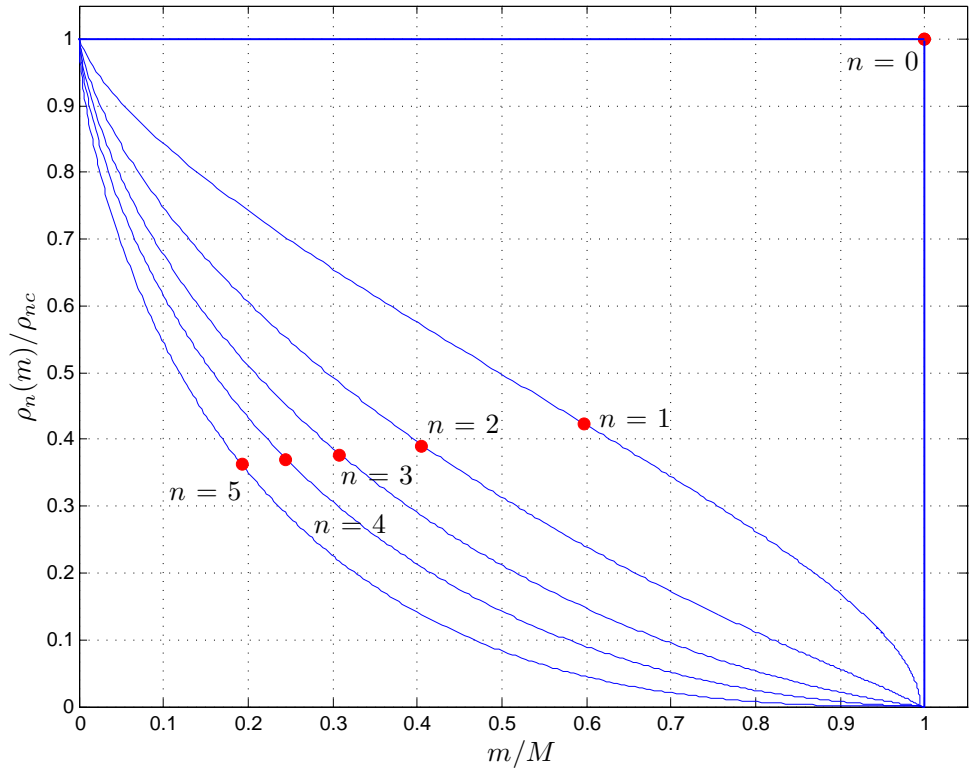}
\caption{Normalized density profiles as a function of fractional included mass $m/M$, for polytropes of mass concentration increasing with $n$. The red dots mark the core radii, at which the densities stay near
$\rho(r_{core})/\rho_c \approx 0.4$, for all $n \geq 1$. For uniformly distributed mass ($n=0$), the polytrope
is all core. As the mass concentration increases ($n\rightarrow 5$), the core shrinks to 20\% of the mass.}
\end{figure}

\begin{description}    %%%%p6b
\item[For $n=0$,] the mass is uniformly distributed, and the entire star is core.
\item[As $0<n<5$ increases,] the fractional core radius shrinks $r_{core}/R\rightarrow 0$, the mass concentrates towards the center, $m_{ncore}/M \rightarrow 0.19$:
    \begin{description}
    \item [For $1<n<3$,] the radius $R$ decreases with mass $M$. Nonrelativistic degenerate stars have $n=3/2$.
    \item [For n=3,] the radius $R$ is independent of mass $M$.  This astrophysically important case will be discussed in Section V and the Appendix.
    \item [For $n>3$,] the radius $R$ increases with mass $M$.
    As $n \rightarrow 5$, the stellar radius increases $\xi_{1n} \rightarrow 3(n+1)/(5-n)$, the core radius shrinks $\xi_{core} \rightarrow \sqrt{10/3 n}$, the fractional core radius $r_{core}/R=\xi_{core}/\xi_{1n} \rightarrow 0.045(5-n)$, $m_{ncore}/M \rightarrow 0.19$, and $_0\omega_n \rightarrow \sqrt{3/\xi_{1n}}\rightarrow 0$.
    \end{description}

\item[For $n=5$,] the stellar radius $R=\infty$ for any mass $M$, while the included mass $m(r)$ is concentrated towards finite $r$. Scaling becomes a variational symmetry, so that the Noether charge $G_5$ in (36) is constant. For the Emden solution, this constant vanishes
    \be G_5 \sim [\xi (\frac{\theta_5'^2}{2}+\frac{\theta_5 ^{6}}{6})+\frac{1}{2}\theta_5 \theta_5 ']=\frac{1}{2}\theta_5\theta_5 '(v_5-u/3-1)=0 \quad, \ee
    so that $v_5=1-u/3,~\theta_5 '=-\frac{\xi \theta_5 ^3}{3}$.
    Integrating then yields
    \be \theta_5(\xi)=(1+\xi^2 /3)^{-1/2}\quad ,\quad \rho_5=\rho_{5c} (1+\xi^2 /3)^{-5/2} \quad,\ee
    after normalizing to $\theta_5 (0)=1$.
\item[For $n>5$] the central density diverges, so that the total mass $M$ would be infinite.
\end{description}

%%%%%%%%%%%%%%%%%%%%%%%%%%%%%%%%%%%%%%%%%%%%%%%%%%%%%%%%%%%%%%%%%%%
\begin{figure}[t] %%%%Fig. 4 Replaces Kennedy's rho-x.eps
\includegraphics[scale=0.75]{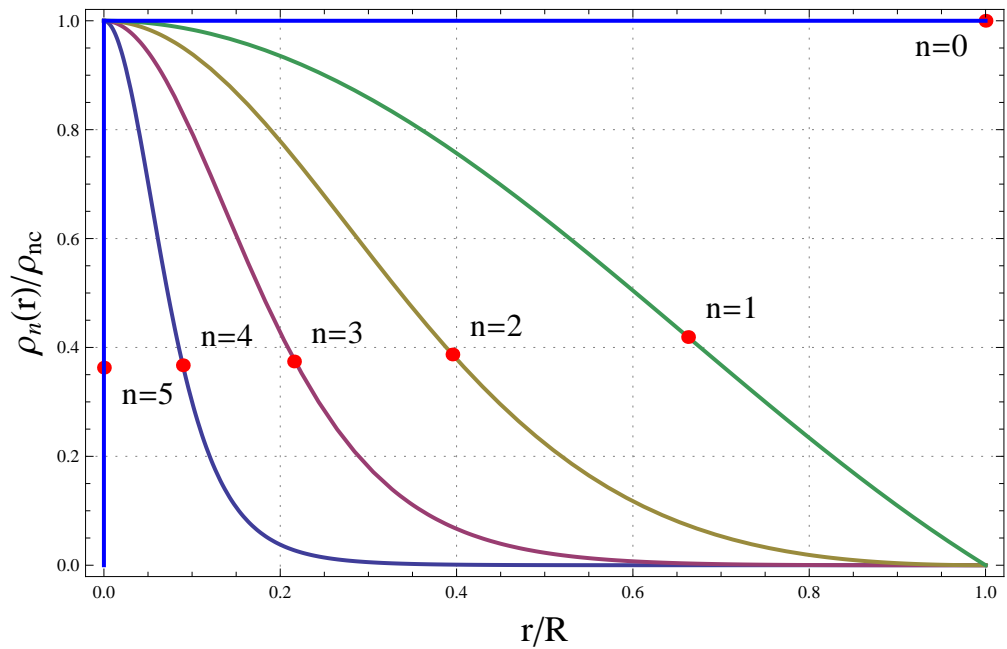}%%Add n=0 horizontal line, n=5 vertical line.
\caption{Normalized density profiles as function of fractional radius $r/R$. The density is uniform for $n=0$, but is maximally concentrated at finite radius for the $n=5$ polytrope, which is unbounded ($R=\infty$). The density at the core radius stays about $ \rho(r_{core})/\rho_c \sim 0.4$, for any $n\geq 1$.}
\end{figure}
%%%%%%%%%%%%%%%%%%

\section{CONCLUSIONS} %%VI p10
We have distinguished two ways in which symmetry reduces a second-order differential equation to a first-order equation:
\begin{description}
\item[Variational Symmetry (epitomized by central field dynamics):] The symmetries of the action lead to conservation laws, first integrals of the original equations of motion
\item[Non-variational Symmetry of the Equations of Motion \\
(epitomized by scale invariant hydrostatics):] Yields a first-order equation between scale invariants which, although not a first integral, still leads to integration by quadratures.
\end{description}
In the latter case, we obtained all the familiar properties of polytropes, {\em directly} from the first-order equation between invariants. We observed that, like all stars, polytropes of index $n$ share a common core density profile and defined a {\em core radius} outside of which their envelopes differ. The Emden functions $\theta_n(\xi)$, solutions of the Lane-Emden equation that are regular at the origin, were finally obtained, along with useful approximations.

The Appendix reviews the astrophysically most important $n=3$ polytrope, describing relativistic white dwarf stars and zero age main sequence stars. While reviewing these well-known applications \cite{Hansen,Kippen}, we stress how these {\em same} mechanical structures differ {\em thermodynamically} and the usefulness of our original (Section V) approximations to these Emden functions.
%%%%%%%%%%%%%%%%%%%%%%%%%%%%%%
\appendix*
\section*{APPENDIX: IMPORTANT ASTROPHYSICAL APPLICATIONS OF $n=3$ POLYTROPES}

The $n=3$ polytrope, which is realized in white dwarfs of nearly maximum mass and in the Eddington standard model for luminous zero-age main sequence (ZAMS) stars that have just started to burn hydrogen, is distinguished by a unique $M-R$ relation: the mass $M=4\pi ~_0\omega_3 ~(K/\pi G)^{3/2}$ is independent of radius $R$, depending on
the constant $K:=P/\rho^{4/3}$.  In these stars, the gravitational and internal energies cancel, making the
total energy $W=\Omega + U=0$, leaving them in dynamical equilibrium at any radius \cite{Kippen,Hansen}.

\subsection{Relativistic Degenerate Stars: $K$ Fixed by Fundamental Constants} %p5

White dwarfs of nearly maximal mass are supported by the degeneracy pressure of relativistic electrons, with number density $n_e=\rho/\mu_e m_H$, where $m_H$ is the atomic mass unit and the number of electrons per atom $\mu_e=Z/A=2$, because these white dwarfs are composed of pure He or $C^{12}/O^{16}$ mixtures. Consequently, $K_{WD}=\frac{hc}{8} [\frac{3}{\pi}]^{1/3} {m_H \mu_e}^{-4/3}$ depends only on fundamental constants. This universal value of $K_{WD}$ leads to the limiting Chandrasekhar mass $M_{Ch}=\frac{\pi^2}{8 \sqrt{15}} M_{\star}/\mu_e^2=5.824 M_{\odot}/\mu_e ^2=1.456 M_{\odot}\cdot(\frac{2}{\mu})^2$ \cite{Hansen,Kippen}.

\subsection{Zero-Age Main Sequence Stars: $K(M)$ Depends on Specific Radiation Entropy} %p5

In an ideal gas supported by both gas pressure $P_{\rm gas}=\mathcal{R}\rho T/\mu :=\beta P$ and radiation pressure $P_{\rm rad}=a T^4 /3 :=(1-\beta) P$, the radiation/gas pressure ratio is
\be \frac{P_{rad}}{P_{gas}}:=\frac{1-\beta}{\beta}=\frac{T^3}{\rho}\cdot\frac{a \mu}{3 \mathcal{R}}\quad.\ee
The specific radiation and ideal monatomic gas entropies are
\be S_{\rm rad}=\frac{4a T^3}{3 \rho}, \quad \quad S_{\rm gas}(r)                                                                    =(\frac{\mathcal{R}}{\mu})\cdot \log{(\frac{T(r) ^{5/2}}{\rho (r)})}\quad  ,\ee
so that the gas entropy gradient
\be \frac{d S_{gas}}{d\log{P}}=(\frac{5\mathcal{R}}{2\mu})\cdot (\nabla -\nabla_{ad})=(\frac{\mathcal{R}}{\mu})\cdot (\frac{\nabla}{\nabla_{ad}} -1) \ee
depends on the difference between the adiabatic gradient $\nabla_{ad}=2/5$ and the star's actual thermal gradient $\nabla:=d\log{T}/d\log{P}$, which depends on the radiation transport.

Bound in a polytrope of order $n$ , the ideal gas thermal gradient and gas entropy gradient are
\be \nabla:=1/(n+1) \quad,\quad \frac{d S_{gas}}{d\log{P}}=(\frac{\mathcal{R}}{\mu})\cdot (\frac{5}{2(n+1)}-1)\quad.\ee
For $n>3/2$, the thermal gradient is subadiabatic, the star's entropy increases outwards, so that the star is stable against convection.

Zero-age main sequence stars (ZAMS), with mass $0.4 M_{\bigodot} <M<150 M_{\bigodot}$, have nearly constant radiation entropy $S_{\rm rad}(M)$, because radiation transport leaves the luminosity generated by interior nuclear burning everywhere proportional
to the local transparency (inverse opacity) $\kappa ^{-1}$.  Assuming constant  $S_{\rm rad} (M)$, we have {\em Eddington's standard model}, an $n=3$ polytrope with $S_{\rm rad}(M)=
4 (\mathcal{R}/\mu)\cdot(1-\beta)/\beta$ and
\be K(M)=P/\rho^{4/3}=\lbrace [3(1-\beta)/a] (\mathcal{R}/\mu\beta )^4 \rbrace^{1/3} ,  \ee
depends only on $\beta (M)$, which is itself determined by
{\em Eddington's quartic equation} \cite{Kippen,Hansen,Chandra}
\be \frac{1-\beta}{\beta ^4}=\Bigl( \frac{M\mu^2}{M_{\star}} \Bigr)^2\quad ,\quad M_{\star}:=\frac{3\sqrt{10} ~_0\omega_3}{\pi ^3} \Bigl( \frac{hc}{G m_H^{4/3}} \Bigr)^{3/2}=18.3 M_{\odot}\quad .\ee

The luminosity $L=L_{Edd} [1-\beta (M)]$ depends on the {\em Eddington luminosity} $L_{Edd}:=4\pi c G M/\kappa_p$ through the photospheric opacity $\kappa_p$. From Eddington's quartic formula, the stellar luminosity
\be L=L_{Edd} (0.003)\mu ^4 \beta  ^4 (M/M_\odot)^3 .\ee
This mass-luminosity relation is confirmed \cite{Hansen} in ZAMS stars: On the lower-mass ZAMS, $\beta\approx 1,~L\sim M^3$; on the upper-mass ZAMS, $\beta\approx \Bigl( \frac{M\mu^2}{M_{\star}} \Bigr)^{-2} \ll 1,~L\sim M $.
\begin{acknowledgments}
Thanks to Dallas Kennedy (The MathWorks) for long collaborations on this topic. Thanks to Andr${\rm\acute{e}}$s E. Guzm${\rm\acute{a}}$n (Universidad de Chile) for calculating the figures with Mathematica and proofreading the manuscript. This work was supported by the Millennium Center for Supernova Science through grant P06-045-F funded by Programa Bicentenario de Ciencia y Tecnolog\'ia de CONICYT and Programa Iniciativa Cient\'ifica Milenio de MIDEPLAN.
\end{acknowledgments}
\bibliography{bibliographyLE}
\end{document}